\documentclass[aip,jcp,floatfix,twocolumn,10pt]{revtex4-1}
\usepackage{graphicx}
\begin{document}
\title{Aggregation of rod-like polyelectrolyte chains in the presence
of monovalent counterions}
\author{Anoop Varghese}
\email{anoop@imsc.res.in}
\author{R. Rajesh}
\email{rrajesh@imsc.res.in}
\author{Satyavani Vemparala}
\email{vani@imsc.res.in}
\affiliation{The Institute of Mathematical Sciences, C.I.T. Campus,
Taramani, Chennai 600113, India}
\date{\today}

\begin{abstract}
Using molecular dynamics simulations, it is demonstrated that monovalent
counterions can induce aggregation of similarly charged rod-like
polyelectrolyte chains.
The critical value of the linear charge density
for aggregation is shown to be close to the critical value for the
extended--collapsed transition of a single flexible polyelectrolyte
chain, and decreases with increasing valency of the counterions. 
The potential of mean force along the center of mass reaction
coordinate  between two similarly charged rod-like
polyelectrolytes is shown to develop an attractive well
for large linear charge densities.  In the attractive regime, the 
the angular distribution of the condensed counterions is no longer
isotropic. 
\end{abstract} 

\maketitle

\section{Introduction}

Polyelectrolytes (PEs) are charged polymers in a solution containing
neutralizing counterions~\cite{dobrynin,netz,levin1}. 
They are common in biological systems, examples include 
DNA~\cite{bloom}, F-actin and microtubules~\cite{tang}. 
A special case is a rod-like PE (RLPE) whose persistence length
is of the order of its contour length. 
RLPEs are of interest theoretically because their thermodynamics is similar to the 
theoretically well studied idealized system of a charged cylinder with 
neutralizing counterions.

PEs, even if similarly charged, may attract each other in the presence
of counterions. The aggregation of similarly charged PEs 
has been extensively studied 
experimentally~\cite{bloom,tang,sedlak,tanahatoe,borsali,ermi,zhang,zribi,bordi,butler}, 
theoretically~\cite{ha3,ha2,ha1,solis,manning2,skhlovskii,potemkin,bruinsma,ermoshkin,pietronave,perico,manning3} 
and
numerically~\cite{jensen,stevens1,diehl,stevens2,allahyarov,savelyev,sayar1,fazli,luan,sayar2}. 
Despite these studies, the role of counterion valency in aggregation 
remains unclear. Though it has been shown unambiguously that multivalent 
counterions can mediate 
aggregation~\cite{bloom,tang,zribi,sedlak,tanahatoe,borsali,ermi,bordi,butler,ha3,ha2,ha1,solis,manning2,skhlovskii,potemkin,bruinsma,ermoshkin,pietronave,perico,jensen,stevens1,diehl,stevens2,allahyarov,sayar1,fazli,luan,sayar2}, 
it is still being debated whether monovalent counterions can cause 
aggregation in the absence of multivalent salts.  There are some  
experimental~\cite{sedlak,tanahatoe,borsali,ermi,zhang} and 
theoretical~\cite{manning2,perico} results that argue for monovalent
counterion induced aggregation of PEs.  At the 
same time, other experimental~\cite{zribi,butler}, 
theoretical~\cite{arenzon1,arenzon2,solis} and 
numerical~\cite{jensen,stevens1,stevens2,diehl,allahyarov,savelyev} studies argue or 
report the absence of aggregation in the presence of monovalent 
counterions.

There are different proposals for understanding the attraction 
between two similarly charged PEs. One of them suggests that the 
attraction is induced by the correlated longitudinal fluctuations of the 
condensed counterion density~\cite{ha1}. This theory predicts 
attraction in the presence of multivalent counterions, 
in agreement with early numerical simulations~\cite{jensen}, for a range of 
system parameters. The 
validity of the Gaussian approximation made in the theory was 
questioned~\cite{levin2,arenzon1}, and more realistic, but simple models 
were considered~\cite{arenzon1,solis}. These models assume localization 
of the condensed counterions on a finite number of allowed sites around 
the PEs, and were able to explain multivalent counterion induced attraction
in terms of spatial distribution of the condensed counterions around the PEs. 
However, these theories predict a very weak~\cite{solis} or
zero~\cite{levin2} attraction  for the case of monovalent counterions.

An alternate approach that explains the attraction of PEs is by Manning et. 
al.~\cite{manning2}, and is based on the classical condensation 
theory~\cite{manning1}. Though this approach does not explicitly 
take into account correlations between counterions, it predicts  
attraction between similarly charged PEs. In fact, the theory predicts 
stronger attraction in the presence of monovalent counterions than 
divalent or trivalent counterions. The origin of attraction is similar to
that of covalent bonds, in that the condensed counterions are shared by the
PEs. This theory was later 
extended~\cite{pietronave,perico}, using extended 
condensation theory~\cite{shurr}, to calculate the interaction free energy
as a function of the separation between the PEs, linear 
charge density of the PEs and valency of the counterions. 
The analysis of the free energy shows that the attraction 
is a consequence of the increase in the number of condensed counterions 
and the counterion condensation volume with decrease in the 
separation between the PEs. This approach differs from the Manning theory
in that the assumption of infinite dilution of added salt is absent. 
Although these theories predict the 
formation of stable aggregates, mediated by monovalent counterions, a clear 
experimental or computational confirmation is still lacking. Most  
earlier numerical
simulations~\cite{jensen,stevens1,stevens2,diehl,allahyarov,savelyev,luan} 
do not observe attraction that is strong enough to form stable
aggregates.

A different phenomena in PE systems, mediated by 
counterions, is the extended--collapsed transition of a single 
flexible~\cite{winkler,brilliantov,solis2,varghese,varghese2,jaya} or 
semi-flexible PE~\cite{golestanian}. As the linear charge density exceeds a 
critical
value, an extended flexible PE chain undergoes a 
first-order transition~\cite{brilliantov,varghese} 
to a globular phase, while an extended semi-flexible PE  transforms into a 
toroid-like phase~\cite{golestanian}. As for aggregation of 
PEs, there are different proposals for the underlying mechanism of the
extended--collapsed transition. In one of the approaches, a
one-component plasma model of the PE system was shown to have negative
compressibility when the linear charge density of the PE chain 
exceeds a critical value~\cite{brilliantov}. The negative
compressibility causes the PE chain to collapse, with the transition 
being first-order.  In an alternate approach~\cite{solis2},
the transition was studied using a two-state model, where the free
energy of the extended and collapsed phases of PE chain was calculated.
The collapsed phase was approximated by an amorphous ionic solid made up
of the condensed counterions and the monomers of the PE chain,
and the strong correlation between the counterions and the monomers
was argued to be responsible for the
collapse of the chain. In an another numerical study~\cite{winkler}, 
the condensed counterions were shown to
form dipoles with the monomers of the chain, and the resultant
dipole-dipole attraction causes the chain to collapse.
A different approach~\cite{golestanian}, 
based on the dependence of the free energy of a
semi-flexible PE chain on counterion density fluctuations, 
argued that the extended phase of a semi-flexible PE chain
destabilizes into a toroid-like phase at high linear charge densities.

In this paper, using molecular dynamics simulations (details in
Sec.~\ref{sec:model}), we demonstrate the 
aggregation of similarly charged RLPEs in the presence of 
monovalent counterions when the linear charge density of the
polymer backbone is higher than a critical value
(Sec.~\ref{sec:aggregation}). We argue that the critical 
backbone charge densities required for the onset of aggregation of
RLPEs and
the extended--collapsed transition of a single flexible PE chain are
nearly equal (Sec.~\ref{sec:collapse}). 
We also
measure the potential of mean force between two RLPEs along the
distance between their centers of mass and show the existence of an
attractive well for large linear charge densities
(Sec.~\ref{effective-interaction}). In this attractive regime, 
a spatial rearrangement of counterions around a RLPE occurs and is
quantified in Sec.~\ref{sec:distribution}. Sec.~\ref{sec:conclusions}
contains a summary and discussion.

\section{Model and simulation method \label{sec:model}}

We model RLPE and flexible PE chains as $N$ spheres (monomers), each 
with charge $+qe$, 
connected through springs. The counterions are modelled as spheres with 
charge $-Zqe$, where $Z=1,2$ and $3$ for monovalent, divalent and 
trivalent counterions respectively. All counterions have same 
valency, and the number of counterions is such that the system is 
overall charge neutral. The polymer chain and the counterions 
are assumed to be in a medium of uniform dielectric constant $\epsilon$.

The interactions between particles $i$ and $j$ are of four types:

Coulomb interaction: The electrostatic energy is given by 
\begin{equation}
U_{c}(r_{ij})=\frac{q_iq_j}{4\pi\epsilon \epsilon_0r_{ij}},
\label{eq.1}
\end{equation}
where $r_{ij}$ is the distance between particle $i$ and $j$, and $q_i$ 
is the charge of the $i^{\text{th}}$ particle, and $\epsilon_0$ is the 
permittivity of free space.

Excluded volume interaction: The excluded volume interactions are 
modelled by the Lennard-Jones potential, which for two particles at a 
distance $r_{ij}$, is given by
\begin{equation}
U_{LJ}(r_{ij})= 4\epsilon_{ij} \left[ 
\left(\frac{\sigma}{r_{ij}}\right)^{12} 
-\left(\frac{\sigma}{r_{ij}}\right)^{6}\right],
\label{eq.2}
\end{equation}
where $\epsilon_{ij}$ is the minimum of the potential and $\sigma$ is 
the inter-particle distance at which the potential becomes zero. We use 
reduced units, in which the energy and length scales are specified in 
units of $\epsilon_{ij}$ and $\sigma$ respectively. The depth of the 
attractive potential $\epsilon_{ij}$ and its range $\sigma$ are set to 
$1.0$ for all pairs of particles. We use shifted Lennard-Jones potential 
in which $U_{LJ}(r_{ij})$ smoothly goes to zero beyond a cut off distance 
$r_{c}$. The value of $r_{c}$ is chosen to be $1.0$ such that the 
excluded volume interaction is purely repulsive for all pairs,
mimicking polymers in good solvents.

Bond stretching interaction: The bond stretching energy for pairs that 
are connected through springs is given by
\begin{equation}
U_b(r_{ij})=\frac{1}{2}k_b(r_{ij}-b)^{2},
\label{eq.3}
\end{equation}
where $k_b$ is the spring constant and $b$ is the equilibrium bond length. 
The values of $k_b$ and $b$ are taken as $500$ and $1.12$ respectively. This 
value of $b$ is close to the minimum of Lennard-Jones potential, and
the spring constant $k_b$ is large enough so that the bond length does not
change appreciably from $b$. 

Bond bending interaction: The rigidity of the polymer backbone is
controlled by a three-body interaction given by 
\begin{equation}
U_\theta(\theta)=k_\theta[1+\cos\theta],
\end{equation}
where $\theta$ is the angle between two adjacent bonds. For RLPEs, we
choose $k_\theta = 10^3$ to ensure rigidity of the backbone, while
$k_\theta$ is set to zero for flexible PEs.

The relative strength of the electrostatic interaction is parameterized 
by a dimensionless quantity $A$:
\begin{equation}
A=\frac{q^{2}\ell_{B}}{b},
\label{eq.4}
\end{equation}
where $\ell_{B}$ is the Bjerrum length~\cite{Russel},
\begin{equation}
\ell_{B}=\frac{e^{2}}{4\pi\epsilon \epsilon_0 k_{B}T},
\label{eq.5}
\end{equation}
where $k_{B}$ is the Boltzmann constant and $T$ is temperature. In our
simulations, we vary $A$ from $0.22$ to $10.93$.

The equations of motion are integrated in time using the molecular 
dynamics simulation package LAMMPS~\cite{lammps1,lammps2}. The 
simulations are carried out at constant temperature (T=1.0), maintained 
through a Nos\'{e}-Hoover thermostat (coupling constant 
$=0.1$)~\cite{nose,hoover}. The system is placed in a cubic box with 
periodic boundary conditions. We use the particle-particle/particle-mesh 
(PPPM) technique~\cite{hockney} to evaluate the energy and forces due to 
the long range Coulomb interactions. The time step for the integration
of the equations of motion is chosen as $0.001$.

In the current study, three systems are considered. In the first
system, aggregation of $50$ similarly charged RLPEs with varying
linear charge densities neutralized by counterions of varying valency
is studied.  In the 
second set of simulations, we locate the extended--collapsed
transition point  of a single flexible PE chain in the presence of
counterions of different valency.
In the third set of simulations, we measure the potential of mean
force between two RLPEs in the presence of monovalent counterions to
understand the observed aggregation. 

\section{Results and Discussion}

\subsection{\label{sec:aggregation} Aggregation of rod-like polyelectrolyte chains}

To study aggregation, we consider a collection of $50$ RLPE chains of 
$30$ monomers each. The system is charge neutralized with monovalent, 
divalent or trivalent counterions. The density of the system is chosen 
to be $4.4\times10^{-4}$ monomers$/\sigma^3$. At this density, the mean
separation between the chains is much larger than the length of the chains. 
We vary the non-dimensional parameter $A$ from $0.22$ to $10.93$.

\begin{figure}
\includegraphics[width=\columnwidth]{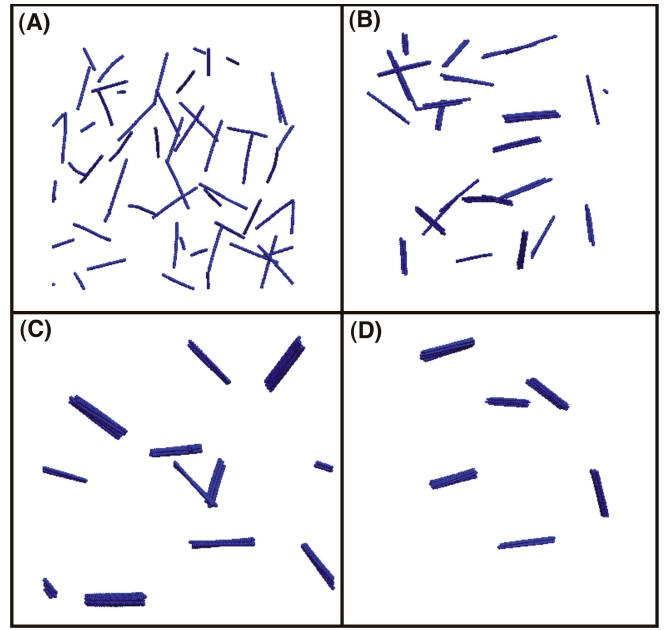}
\caption{The two dimensional projection of the RLPEs 
at time steps (A)~$0$, (B) $1.90\times10^6$, (C) $8.90\times10^6$ and 
(D) $4.72\times10^7$. The counterions are monovalent and are 
not shown for clarity. The RLPEs aggregate, with the number of
aggregates decreasing in time. The data are for $A=9.43$.}
\label{fig2}
\end{figure}

We first show that monovalent counterions can mediate aggregation of 
similarly charged RLPEs, in contrast to the previous 
experimental~\cite{zribi,butler}, 
theoretical~\cite{arenzon1,arenzon2,solis} and 
numerical~\cite{jensen,stevens1,stevens2,diehl,allahyarov,savelyev} studies. In 
Fig.~\ref{fig2}(A)--(D), we show snap shots of the system 
at increasing times for $A=9.43$, when all counterions are monovalent. 
For this value of $A$, electrostatic interactions are much larger than thermal 
energies, and aggregation of RLPEs can clearly be seen.

We quantify aggregation by calculating the dependence of 
average aggregate size on $A$.
Two PE chains are defined to form an aggregate if the 
distance between any monomer of the first chain and any monomer of the 
second chain is less than $2\sigma$. Similarly, a PE chain is defined to 
be part of an aggregate of size $m$ ($m>2$) if the distance between
any of its monomers and any monomer of the other $m-1$ chains 
is less than $2\sigma$. Other similar definitions for what 
constitutes an aggregate can be found in the 
literature~\cite{stevens1,sayar1,sayar2}, but we find that the results 
do not change noticeably with different choices of the definition.

\begin{figure}
\includegraphics[width=0.85\columnwidth]{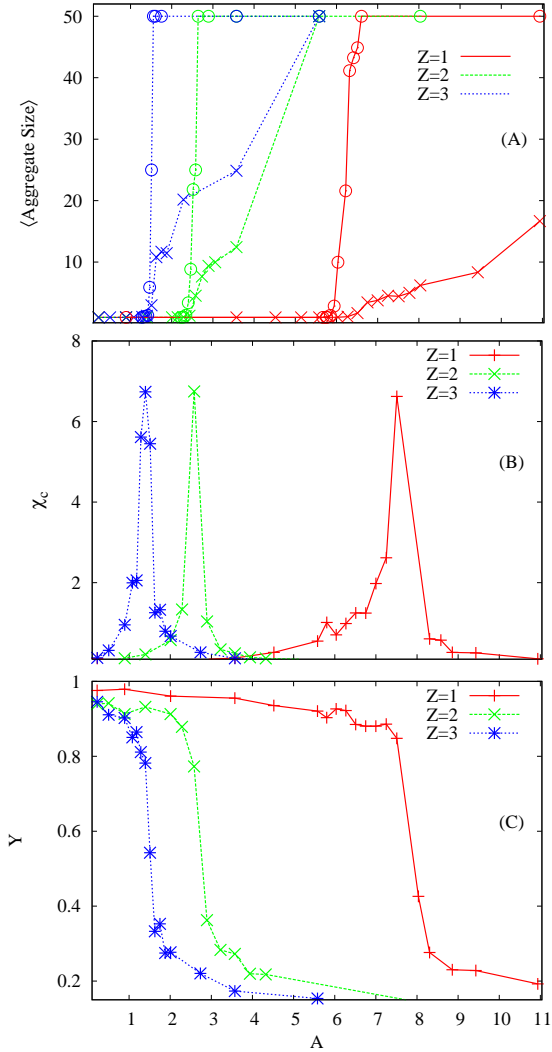}
\caption{ (A) The variation of the average aggregate size with $A$.
The data shown by crosses are obtained from simulations where the
RLPEs are initially randomly distributed. For the data shown by
circles we start at high value of $A$ where initially all RLPEs are
part of a single aggregate. The initial configuration for a lower
value of $A$ is the final configuration of the previous value of $A$.
(B) The relative fluctuation $\chi_c$ in electrostatic energy of a
single flexible PE as defined in 
Eq.~(\ref{eq.10}). The height of the peaks have been rescaled for clarity. 
(C) The variation of asphericity $Y$ [see Eq.~(\ref{eq.8})] with $A$
of a single flexible PE chain.}
\label{fig1}
\end{figure}

Let $N_m$ be the number of aggregates of size $m$. Then, the average 
aggregate size is given by $\sum_{m=1}^{50} m N_m/ \sum_{m=1}^{50} N_m$. 
Measuring the equilibrium value of $N_m$ turns out to be difficult in
the aggregate regime because of large equilibration times and
limited computational time. We, therefore, measure $N_m$ after
starting from two different types of initial conditions and discarding
a fixed number of initial simulation steps. In the first type, we
start from an initial random configuration of RLPEs and discard the
first $3\times 10^7$ time steps. We then average the aggregate size over
the next $10^7$ time steps. The average aggregate size, thus computed, 
is shown by crosses in Fig.~\ref{fig1}(A) as a function of $A$ for
counterions of different valency. 
The aggregation begins at critical values 
$A=6.57, 2.50$ and $1.45$ for monovalent, divalent 
and trivalent counterions respectively, which 
correspond to an average aggregate size of $2.0$. At high enough values of $A$, 
all the $50$ chains form a single aggregate for both divalent 
and trivalent counterions. However, in the case of monovalent counterions, 
even for the highest 
value of $A$ (10.93) that we studied, we did not observe formation of
a single aggregate within the maximum
duration of our simulations ($5\times10^7$), and the final configuration of 
the system consists three aggregates of approximately $16$ PEs each. 
In the second set of simulations, we start with a high value of $A$
and an initial condition where all the chains are in an aggregate. We
then  discard the first $10^7$ time steps, and measure the aggregate
size over the next $10^6$ time steps. $A$ is then decreased and the
initial condition is chosen to be final state of the previous $A$
value. The data thus obtained are shown by circles in
Fig.~\ref{fig1}(A). The values for $A$ for which the two kinds of runs
give the same value for average aggregate size can be taken to be the
correct equilibrium value. The formation of a hysteresis loop is
clearly seen.

\begin{figure}
\includegraphics[width=\columnwidth]{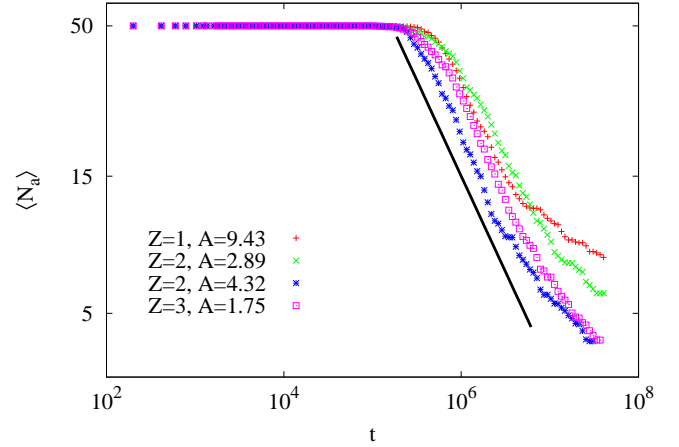}
\caption{The variation of $\langle N_a \rangle$, the average number of 
aggregates,  with time $t$ for different values of $A$ and valency. 
The data is log-binned and averaged over five different initial
configurations. The solid black line is a power law $\sim t^{-2/3}$}
\label{kinetics}
\end{figure}

The difficulty in equilibration in the aggregating regime is due to
large diffusion time scales. As small aggregates form, they take
longer and longer to diffuse and come close to each other before the
next aggregation event can occur.  We find that for values of $A$ 
larger than the critical values, the aggregation kinetics is 
independent of $A$ and the valency of the counterions. In
Fig.~\ref{kinetics}, we show the time evolution of the history
averaged number of aggregates  $\langle N_a \rangle$. 
At the start of the simulations, the PEs are randomly distributed 
$(\langle N_a \rangle =50)$. After initial transients, corresponding
to counterion condensation, $\langle N_a \rangle$ decays as a power
law $t^{-\tau}$, with $\tau \approx 0.66$, independent of valency. If
we model aggregation as irreversible coalescence of point-sized
particles (see Ref.~\cite{leyvraz2005,ccareview} for a review), 
then the above power law decay corresponds to the 
diffusion constant of an aggregate of size $m$ being proportional to 
$m^{-1/2}$.

In the earlier simulations~\cite{jensen,stevens1,stevens2,diehl,allahyarov,savelyev}, the interaction
between similarly charged RLPEs in the presence of monovalent
counterions was found to be repulsive. Typical values of $A$ used in
these simulations are $2.10$~\cite{jensen},
$2.90$~\cite{stevens1,stevens2} and $4.17$~\cite{diehl}, which are
much smaller than the critical value $(\approx6.57)$ above which we
observe the aggregate formation. We suggest that the absence of
attractive interactions required for aggregation in these simulations
can be attributed to the inappropriate values of $A$.

\subsection{\label{sec:collapse} Extended-Collapsed transition of a single flexible polyelectrolyte 
chain}

In our earlier simulations~\cite{varghese,varghese2}, we had demonstrated 
that a 
single flexible PE chain undergoes a first-order transition from an 
extended to a collapsed configuration, mediated by the 
counterions, when the parameter $A$ exceeds a certain critical value. To 
establish a possible relation between the aggregation phenomena of 
RLPE chains and the extended--collapsed transition of a single 
flexible PE chain, we study the extended--collapsed transition of a single 
flexible PE chain of $600$ monomers in the presence of monovalent and 
multivalent counterions. The density of the system is chosen as 
$2.7\times10^{-6}$ monomers$/\sigma^3$ such that the direct contact 
between the PE chain and it's periodic images is avoided. The initial 
configuration of the PE chain is randomly chosen and the counterions are 
uniformly distributed inside the simulation box. The system is allowed 
to equilibrate for $10^7$ steps and the averages are taken over a 
production run of $10^7$ steps.

A useful quantity to study the extended--collapsed transition is the 
electrostatic energy per monomer $E_c$. $E_c$ shows different scaling 
with the number of monomers in the extended and collapsed 
phases~\cite{varghese}. The relative fluctuation $\chi_c$in $E_c$ is 
defined as
\begin{equation}
\chi_c=\frac{N\left[\langle E_c^{2}\rangle-\langle E_c
\rangle^{2}\right]}{\langle E_c \rangle^{2}}.
\label{eq.10}
\end{equation}
Fig.~\ref{fig1}(B) shows the variation of $\chi_c$ with $A$. $\chi_c$ 
has a peak around $7.5, 2.57$ and $1.40$ for monovalent, divalent and 
trivalent counterions respectively. These peaks correspond to the 
extended--collapsed transition of the single PE chain.

The extended--collapsed transition can be further quantified by studying 
the variation of asphericity of the PE chain as a function of $A$. We 
define asphericity as
\begin{equation}
Y=\left\langle\frac{\lambda_{1} - \frac{\lambda_{2} + \lambda_{3}}{2}} 
{\lambda_{1}+\lambda_{2} + \lambda_{3}}\right\rangle,
\label{eq.8}
\end{equation}
where $\lambda_{1,2,3}$ are the eigenvalues of the moment of inertia 
tensor with $\lambda_{1}$ being the largest eigenvalue. The moment of 
inertia tensor $G$ is
\begin{equation}
G_{\alpha\beta}=\frac{1}{N}\sum_{i=1}^{N}r_{i\alpha}r_{i\beta},
\label{eq.9}
\end{equation}
where $r_{i\alpha}$ is the $\alpha^{th}$ component of the position 
vector $\vec{r}_{i}$. Asphericity $Y$ is zero for a sphere (collapsed 
globule) and one for a linear rod (extended configuration). For all 
other configurations, it has a value between zero and one.

The variation of $Y$ with $A$ is shown in Fig.~\ref{fig1}(C). For all 
the three valencies, $Y$ makes a transition from a value close to one 
(extended configuration) to zero (collapsed configuration) at a critical 
value of $A$. The transition in $Y$ occurs at the same critical value of $A$ 
where $\chi_c$ has a peak, and corresponds to the extended--collapsed
transition of the PE chain. 
We note that these critical values roughly coincide with the critical 
values for the aggregation of RLPE chains in the presence 
counterions of the corresponding valency. This suggests the
possibility that the 
underlying mechanisms of the collapse of a single PE chain and the 
aggregation of RLPE  chains are closely related.

\subsection{\label{effective-interaction}The effective interaction potential 
between two rod-like PE chains with monovalent counterions}

To understand the observed aggregation of RLPEs in the presence of
monovalent counterions, we measure the effective interaction potential 
between two similarly charged RLPEs.
The effective interaction potential, for a separation $d$ 
between the rods, is equal to the reversible work $W(d)$ done in 
bringing them from infinite separation to a separation of $d$. Since the 
system is overall charge neutral, we expect the interaction potential to 
be short-ranged, and hence $W(d)$ can be approximated by the work done 
in bringing the rods to a separation $d$ from a finite separation $d'$, 
where $d'>d$. $W(d)$ is then $\int_{d'}^{d}dxf(x)$, where $f(x)$ is 
the normal component of the force required to keep the two rods at a 
separation $x$.
\begin{figure}
\includegraphics[width=\columnwidth]{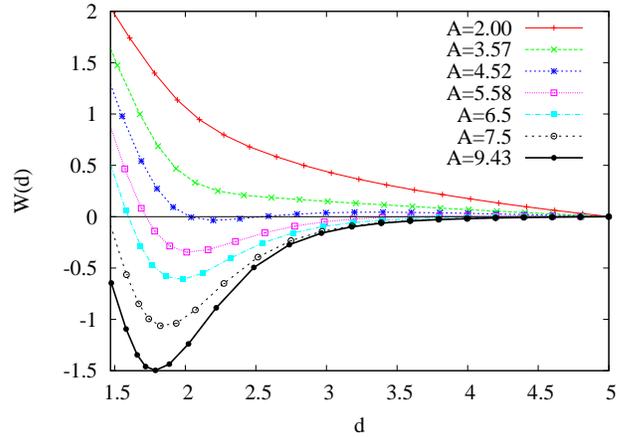}
\caption{The reversible work $W$ needed to bring two rod-like PEs from a
separation of $5\sigma$ to $d$, in the presence of monovalent
counterions. 
The averages are taken over a production run of
$10^7$ steps, after the system  has been equilibrated over $10^7$ steps.}
\label{fig3}
\end{figure}

For the evaluation of $W(d)$, we consider two parallel RLPEs of $30$ 
monomers each. The system is charge neutralized by monovalent 
counterions. The density of the system is chosen to be 
$4.4\times10^{-4}$ monomers$/\sigma^3$. Fig.~\ref{fig3} shows the 
variation of $W(d)$ per monomer for different values of $A$. For low 
values of $A$, $W(d)$ is always positive and decreasing with $d$, 
showing repulsion between the rods. At high values of $A$, $W(d)$ 
develops a minimum which is the onset of attraction between the rods.
When the depth of the potential s comparable to thermal energy $T$, 
the attraction will
be strong enough to form stable aggregates.

A similar evaluation of $W(d)$ for two similarly PEs in the presence of 
monovalent counterions was carried out in earlier 
simulations~\cite{stevens2,allahyarov}. These simulations were performed 
at a range of values of $A$ which is much less than the values at which 
we observe the attractive part in $W(d)$, and these simulations failed 
to observe the attractive part in $W(d)$. In Ref.~\cite{solis}, solving 
a simple model of PE system with monovalent counterions,  
the free energy was evaluated as a function of the separation between
the rods.  For the value of $A$ studied ($A=4.10$), it was shown that
the depth of the attractive well was insufficient to bind the two PEs.

\subsection{\label{sec:distribution} The angular distribution of the condensed counterions}

It has been observed in experiments~\cite{angelini} and 
simulations~\cite{deserno} that the angular distribution of the 
condensed multivalent counterions changes as the PEs approach each 
other. Many theoretical studies~\cite{levin2,arenzon1,solis} also argue 
that the spatial arrangement of the condensed multivalent counterions 
around the PEs plays an important role in developing an attractive 
interaction between the PEs.

\begin{figure}
\includegraphics[width=\columnwidth]{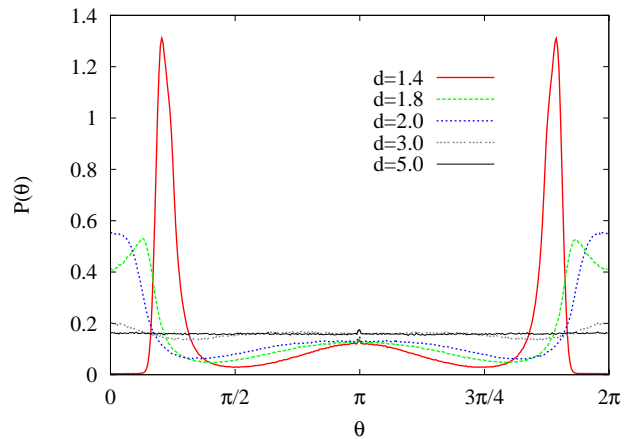}
\caption{The angular distribution $P(\theta)$
of condensed monovalent counterions around a given PE for $A=9.43$.}
\label{fig4}
\end{figure}

We study the relationship between spatial distribution of counterion and 
the effective interaction between the RLPEs when all counterions are 
monovalent. For this,  we consider the same 
system specifications as in Sec.~\ref{effective-interaction}. A 
counterion is said to have condensed on a PE if its separation from any 
of the monomers of the PE is less than $1.25\sigma$. When the PEs come 
close by, the counterion may be shared by both the PEs, but the 
qualitative nature of the angular distribution remains the same. In 
Fig.~\ref{fig4}, we show the dependence of $P(\theta)$ on $\theta$, 
where $P(\theta) d\theta$ is the probability that the condensed 
counterion is between angle $\theta$ and $\theta+d \theta$. 
The data shown are for $A=9.43$, for which a 
pronounced minimum in $W(d)$ was observed (see Fig.~\ref{fig3}). 
Here, $\theta=0$ or $2 \pi$ corresponds
to a location in the same plane and in  between the two PEs,  
while $\theta=\pi$ 
corresponds to the counterion being located in the same plane but away
away from the second PE. 
When the separation $d$  between the PEs is much larger than $1.85 \sigma$,
the condensed  counterions are distributed uniformly around the PE. As the PEs approach 
each other, the distribution develops a sharp peak at $\theta=0$, and a 
broad peak at $\theta=\pi$, showing that most condensed counterions are 
located in the same plane as that of the PEs, and mostly between the 
PEs. For separations less than $1.85\sigma$, the peak at $\theta=0$ is 
shifted to higher values, showing that the condensed counterions are 
expelled out of the plane of the two PEs, though they remain in the 
region between the PEs. This result is in accordance with Manning's
theory~\cite{manning2} of shared counterions being the origin of
attraction between two similarly charged PEs. 
The exclusion of counterions unscreens the 
interaction between the similarly charged monomers of the two PEs, 
causing the repulsion between them. These observations are consistent 
with the nature of the effective interaction potential between the PEs, 
i.e., attraction for intermediate separation($1.85\sigma \lesssim d 
\lesssim4\sigma$) and repulsion at very small separation 
($d\lesssim1.85\sigma$).

\section{Conclusions \label{sec:conclusions}}

We studied, using molecular dynamics simulations, the role of the 
valency of the counterions on the aggregation of similarly charged 
RLPE chains. We showed that monovalent counterions 
can mediate the aggregation, provided the linear charge 
density of the PE backbone is larger than a critical
value. The absence of aggregation in the presence of monovalent counterions 
in earlier simulations and experiments is probably due to the linear
charge density being smaller than the critical value.
The critical linear charge density for aggregation decreases with increasing counterion valency,
and is found to be close to the critical value for the extended--collapsed transition of a single
flexible PE chain. We also find that, for two parallel
RLPEs, the angular distribution of the condensed counterions
changes with the separation between the chains. When the effective
interactions are attractive, we find that the condensed counterions
are shared by the RLPEs, similar to the mechanism suggested by
Manning~\cite{manning2}. 

It would be interesting to study the aggregation transition in more
detail. From our simulations, it appears that the phase transition is
first-order in nature. However, using molecular dynamics simulations,
it is difficult to equilibrate the system in the aggregating regime,
especially if the initial density is low. Hybrid simulations using both
Monte Carlo and molecular dynamics might be useful to study the 
transition~\cite{sayar1,sayar2}.

Another problem of interest is the kinetics of aggregation. From our
simulations, it appears that the density decay is independent of $A$
and valency of the counterions. This may be due to fact that the
attraction between two RLPEs is short-ranged for all values of $A$ and 
valency. When two aggregates come close by,
the probability that they aggregate may depend on $A$ and valency.
This, while affecting the prefactor of the power law decay, does not affect
the exponent, suggesting that the kinetics is diffusion limited. 
Understanding the kinetics better, and also its dependence on
hydrodynamic interactions will be part of a future study.

We note that the critical linear charge density for aggregation of RLPEs 
in the
presence of monovalent counterions is much higher than that in the presence
of multivalent counterions. One system which has comparable high
linear charge densities is solutions of charged worm-like micelles
\cite{micelleexpt1,micelleexpt2}. Earlier Monte Carlo simulations of such
systems have used high charge densities \cite{micellmc}. Monovalent
counterion induced aggregation, as seen in the current paper, may be realized
in such charged micellar systems.

\begin{acknowledgements}
All simulations were carried out on the Intel Nehalem 2.93 GHz supercomputing 
machine Annapurna at The Institute of Mathematical Sciences.
\end{acknowledgements}


%

\end{document}